\documentclass[letterpaper,twocolumn,10pt]{article}
\usepackage{usenix2019,epsfig,endnotes}

\usepackage{silence}
\usepackage[breaklinks,colorlinks]{hyperref}
\usepackage[usenames,dvipsnames]{xcolor}
\hypersetup{citecolor=blue,linkcolor=blue}
\usepackage{cite}
\usepackage{amsmath,amsopn,amssymb}
\usepackage{endnotes,microtype,xspace,graphicx,fancyvrb,multirow}
\usepackage{booktabs}
\usepackage{array,underscore,relsize}
\usepackage[T1]{fontenc}
\usepackage{times}
\usepackage{fancyhdr,lastpage}
\usepackage{fancyhdr}
\usepackage{enumitem}
\usepackage{subcaption}
\usepackage[labelfont=bf,font=small,skip=5pt]{caption}
\usepackage{fp}

\usepackage{balance}

\newcommand{\sys}{\mbox{\cc{libmpk}}\xspace}

\newcommand{\cc}[1]{\mbox{\smaller[0.5]\texttt{#1}}}

\fvset{fontsize=\scriptsize,xleftmargin=8pt,numbers=left,numbersep=5pt}

\input{code/fmt}
\newcommand{\figrule}{\hrule width \hsize height .33pt}
\newcommand{\coderule}{\vspace{-0.5em}\figrule\vspace{0em}}

\def\Snospace~{\S{}}

\newif\ifdraft\drafttrue
\newif\ifnotes\notestrue
\ifdraft\else\notesfalse\fi

\input{glyphtounicode}
\newcolumntype{R}[1]{>{\raggedleft\let\newline\\\arraybackslash\hspace{0pt}}p{#1}}

\newcommand{\squishlist}{
\begin{itemize}[noitemsep,nolistsep]
  \setlength{\itemsep}{-0pt}
}
\newcommand{\squishend}{
  \end{itemize}
}

\usepackage{tikz}
\newcommand*\circled[1]{\tikz[baseline=(char.base)]{\node[shape=circle,fill,inner sep=0.5pt] (char) {\textcolor{white}{#1}};}}

\newcommand*\BC[1]{%
\begin{tikzpicture}[baseline=(C.base)]
\node[draw,circle,fill=black,inner sep=0.2pt](C) {\textcolor{white}{#1}};
\end{tikzpicture}
}

\usepackage{xstring}
\newcommand{\PP}[1]{
\vspace{1px}
\noindent{\bf \IfEndWith{#1}{.}{#1}{#1.}}
}

\newcommand{\PPNoCheck}[1]{
\vspace{1px}
\noindent{\bf {#1.}}
}

\newcommand{\X}{{\footnotesize $\times$}\xspace}

\newcommand{\KB}{\,\text{KB}\xspace}

\newcommand{\GB}{\,\text{GB}\xspace}

\newcommand{\pkru}{\cc{PKRU}\xspace}
\newcommand{\wrpkru}{\cc{WRPKRU}\xspace}
\newcommand{\rdpkru}{\cc{RDPKRU}\xspace}
\newcommand{\eax}{\cc{EAX}\xspace}
\newcommand{\ecx}{\cc{ECX}\xspace}
\newcommand{\edx}{\cc{EDX}\xspace}
\newcommand{\pkfree}{\cc{pkey\_free()}\xspace}
\newcommand{\pkalloc}{\cc{pkey\_alloc()}\xspace}
\newcommand{\pkmprotect}{\cc{pkey\_mprotect()}\xspace}
\newcommand{\mpkprotect}{\cc{mpk\_mprotect()}\xspace}
\newcommand{\pkeysync}{\cc{do\_pkey\_sync()}\xspace}
\newcommand{\mprotect}{\cc{mprotect()}\xspace}
\newcommand{\domprotect}{\cc{do\_mprotect\_pkey()}\xspace}

\newcommand{\mpk}{MPK\xspace}
\newcommand{\mpkalloc}{\cc{mpk\_malloc()}\xspace}
\newcommand{\mpkfree}{\cc{mpk\_free()}\xspace}
\newcommand{\mpkbegin}{\cc{mpk\_begin()}\xspace}
\newcommand{\mpkend}{\cc{mpk\_end()}\xspace}
\newcommand{\mpkinit}{\cc{mpk\_init()}\xspace}
\newcommand{\mpkmmap}{\cc{mpk\_mmap()}\xspace}
\newcommand{\mpkdestroy}{\cc{mpk\_munmap()}\xspace}
\newcommand{\mmap}{\cc{mmap()}\xspace}
\newcommand{\munmap}{\cc{munmap()}\xspace}

\newcommand{\xmm}{\cc{xmm}\xspace}
\newcommand{\rbx}{\cc{rbx}\xspace}
\newcommand{\rdx}{\cc{rdx}\xspace}

\newcommand{\ie}{i.e.}

\newcommand{\boxbeg}{
\noindent
\vspace{2px}
\begin{tabular}{|l|}\hline
\begin{minipage}{0.94\columnwidth}
\vspace{2px}
\noindent
}

\newcommand{\boxend}{
\vspace{1px}
\end{minipage}\\ \hline
\end{tabular}
\vspace{1px}
}

\gdef\therev{dd08f74}
\gdef\thedate{2018-11-15 13:25:07 -0800}

\usepackage{titling}
\usepackage[compact]{titlesec}

\begin{document}
\title{\sys: Software Abstraction for Intel Memory Protection Keys \vspace{-0.5em}}

\ifdefined\DRAFT
 \pagestyle{fancyplain}
 \lhead{Rev.~\therev}
 \rhead{\thedate}
 \cfoot{\thepage\ of \pageref{LastPage}}
\fi

\author{
  Soyeon Park\;
  Sangho Lee$^\dagger$\;
  Wen Xu\;
  Hyungon Moon$^\ast$\;
  Taesoo Kim\;
  \\\\
  \emph{Georgia Institute of Technology} \\
  \emph{$^\dagger$Microsoft Research} \\
  \emph{$^\ast$Ulsan National Institute of Science and Technology}
  }

\date{}
\maketitle

\begin{abstract}
Intel memory protection keys (MPK) is a new hardware feature
to support thread-local permission control on groups of pages
without requiring modification of page tables.
Unfortunately, its current hardware implementation and
software supports suffer from
security, scalability, and semantic-gap problems:
(1) MPK is vulnerable to protection-key-use-after-free and protection-key corruption;
(2) MPK does not scale due to hardware limitations; and
%
(3) MPK is not perfectly compatible with \mprotect
because it does not support permission synchronization across threads.

In this paper, we propose \sys,
a software abstraction for MPK.
\sys virtualizes protection keys to
eliminate the protection-key-use-after-free and protection-key corruption problems
while supporting a tremendous number of memory page groups.
\sys also prevents
unauthorized writes to its metadata and supports inter-thread
key synchronization.
%
%
We apply \sys to three
real-world applications: OpenSSL, JavaScript JIT
compiler, and Memcached for memory protection and isolation. 
An evaluation shows that \sys introduces
negligible performance overhead (<1\%) compared with insecure
versions, and improves their performance by 8.1\X
over secure equivalents using \cc{mprotect()}.
The source code of \sys will be publicly available
and maintained as an open source project.

\end{abstract}

\section{Introduction}
\label{s:intro}

Operating systems (OSs) rely on the memory management unit (MMU)
to define and enforce a process's access right
to a memory page.
The page table maintains
a page table entry (PTE) for each page
specifying the permission,
e.g., readable, writable, or executable,
and the MMU checks it
to determine the legitimacy of each memory access.
OSs can change the permission
by updating PTEs and
flushing the translation lookaside buffer (TLB)
to reload the updated PTEs into the TLB.

Alternatively,
some instruction set architectures (ISAs),
e.g., ARM~\cite{arm} and IBM Power~\cite{power},
allow OSs
to assign a \textit{key} to each page,
and define the access rights of a process using
a CPU register.
This effectively classifies the memory into multiple
groups with the same access rights.
%
%
Changing a process's access right to a group of pages
is as easy as updating a CPU register,
which commonly takes about tens of cycles at most.

Maintaining page access rights is important
to prevent attackers from
accessing and manipulating arbitrary memory locations
to eliminate information leakage and memory corruption attacks.
%
An arbitrary read vulnerability of an application
can lead 
to leak the application's sensitive information kept in the memory.
For example,
the Heartbleed vulnerability of the OpenSSL library~\cite{heartbleed}
allows attackers to
steal the sensitive data of applications using the library,
including cryptographic private keys and passwords.
Further, protecting control data from the memory corruption attacks
helps to prevent ways to achieve arbitrary code execution.
%
By using information leakage vulnerabilities,
attackers aim to
find the locations of control data
(e.g., return addresses and virtual function tables) and
corrupt them with arbitrary write vulnerabilities,
to carry out control-flow hijacking attacks~\cite{jitrop}.
%

A number of software- or hardware-based mechanisms
have been proposed to
ensure the confidentiality and integrity of the memory.
Some of them~\cite{watchdog2012,oscar,softbound} are designed
to prevent a class of attacks completely.
However, due to their high costs,
others~\cite{chen:shreds16,song:hdfi,guan:trustshadow, mogosanu:microstache, frassetto:imix} focus on minimizing the impact of
single vulnerability by isolating
a small, critical portion of code and data only,
which are practical but have coverage and/or scalability problems.
%

Recently, Intel deployed
a hardware feature,
known as \emph{memory protection keys} (\mpk).
\mpk has three advantages over page-table-based mechanisms:
%
(1) \emph{performance},
(2) \emph{group-wise control}, and
(3) \emph{per-thread view}.
First,
\mpk utilizes
a \emph{protection key rights register} (\pkru)
to maintain
the access rights of individual protection keys
associated with specific memory pages:
read/write, read-only, or no access.
Unlike the page-table-based mechanisms that
flush the TLB and
update kernel-level page-table data structures (virtual memory areas, VMAs)
to change access rights,
taking more than 1,000 cycles,
\mpk only requires to execute a non-privileged instruction \wrpkru to
update \pkru,
taking around 20 cycles (\autoref{s:measure}).
In addition,
\mpk enables \emph{execute-only memory}~\cite{lie:xom00}
because its access rights are orthogonal to
whether the page is executable.

Second,
\mpk supports up to 16 different protection keys;
that is,
there can be up to 16 different \emph{memory page groups}
consisting of memory pages with the same protection keys.%
~\footnote{The default group (0) is special, so only 15 groups are effective in general.}
The pages that compose a group
do not need to be contiguous, and
their access right can be changed at once.
This group-wise control allows developers to
change access rights to memory page groups
in a fine-grained manner according to
the types and contexts of data stored in them.
%
For example,
a server application can associate
different memory page groups with different sessions or clients
to protect them individually.

Third,
\mpk allows each thread (i.e., each hyperthread)
to have a unique \pkru,
realizing per-thread memory view.
Even if two threads share the same address space,
their access rights to the same page can differ
(e.g., read/write versus read-only).
%

Unfortunately,
we find that
the current hardware implementation of Intel \mpk and
its standard library and kernel supports
suffer from
(1) \emph{security},
(2) \emph{scalability}, and
(3) \emph{semantic-gap} problems,
hindering its widespread adoption.
First,
\mpk suffers from
\emph{protection-key-use-after-free} and
\emph{protection key corruption} problems.
The protection key assigned to a page group
can be re-used
after deallocation via \pkfree system call.
%
However,
\pkfree does not
initialize the protection key field of the page group
associated with the deallocated key,
resulting in ambiguity when
the deallocated key is
re-assigned to a different group using \pkalloc.
Also,
if attackers know an arbitrary write vulnerability,
they can corrupt \mpk protection keys 
stored in a variable
to manipulate an application
to change the permission of a target page group.
%

Second,
\mpk does not scale because
its \pkru can manage up to 16 protection keys
due to a hardware restriction.
When an application tries to
allocate more than 16 protection keys,
\pkalloc just returns an error,
implying that the application itself
should implement its own mechanism
if it has to deal with more than 16 memory page groups.

Third,
\mpk has a semantic gap
with the conventional \mprotect
because, unlike \mprotect working at a process level,
\mpk works at a thread level,
which results in potential security and performance problems.
For example,
Linux kernel implements
an execute-only memory feature with \mpk:
\cc{mprotect(addr, len, PROT_EXEC)}.
However,
this \mpk-based execute-only memory
does not consider inter-thread permission synchronization,
which should be ensured by \mprotect by nature.
This allows
another thread that did not execute \mprotect
to have a chance to read the execute-only memory.
%
Further,
since some applications still assume
a process-level memory permission model,
they cannot benefit from MPK's efficiency
and group-wise control, 
unless they synchronize access rights
across threads by themselves.
%

In this paper,
we propose \sys,
a secure, scalable, and semantic-gap-mitigated
software abstraction to
fully utilize \mpk
in a practical manner.
%
\sys implements
(1) \emph{protection key virtualization}
to eliminate the protection-key-use-after-free problem and
to support a virtually infinite number of memory page groups,
(2) \emph{metadata protection}
to prevent attackers from tricking \mpk
by corrupting the protections keys in the memory,
and
(3) \emph{inter-thread key synchronization}
to ensure the semantics of \mprotect with \mpk.

First,
\sys provides \emph{virtual protection keys}
to applications
to hide real hardware keys from them.
This design avoids the protection-key-use-after-free problem
by scheduling the mappings between
virtual protection keys and hardware protection keys.
With the virtual key scheduling,
\sys scales \mpk to
support a virtually infinite number of protection keys
with the same semantics:
group-wise control and per-thread view.

Second,
\sys protects its metadata from corruption.
Basically,
\sys makes all virtual protection keys read-only
by hardcoding them to the code and
enforcing direct calls.
%
%
%
All other important metadata,
including the mappings between virtual and hardware keys,
are maintained in the kernel
to avoid corruption
while avoiding unnecessary system calls
to minimize the performance overhead.
In addition, \sys is designed across the layers---the kernel and user---%
to protect its internal metadata from malicious overwrites through
privilege separation,
and yet to maximize the performance cost by avoiding unnecessary
system calls.

Third,
\sys provides
an efficient inter-thread key synchronization technique
to utilize \mpk
as an efficient alternative of \mprotect
with the same semantics.
%
It is 1.7\X--3.8\X faster than \mprotect
while varying the number of 4KB pages from 1 to 1,000 (\autoref{s:micro}).
%
This huge performance improvement
benefits from our lazy \pkru synchronization technique and
lacks of TLB flush and VMA update.
%

To show the effectiveness and practicality of \sys,
we apply it to three real-world applications:
OpenSSL library, JavaScript just-in-time (JIT) compiler, and Memcached.
First,
we modify the OpenSSL library
to create secure memory pages
for storing cryptographic keys
to mitigate information leakage.
%
%
Second,
we modify three JavaScript JIT compilers
(SpiderMonkey, ChakraCore, and v8)
to protect the code cache
from  memory corruption,
by enforcing the W$\oplus$X security policy.
%
%
Third,
we modify Memcached to secure
almost all its data, including
slab and hash table
whose size can be several gigabytes.
%
The evaluation results show that
\sys and its applications
have negligible overhead (<1\%).


We summarize the contributions of this paper as follows:

\squishlist
\item \textbf{Comprehensive study.}
  We study the design, functionality, and characteristics
  of Intel \mpk in detail.
  Further,
  we identify the critical challenges of utilizing
  \mpk:
  security (protection-key-use-after-free and protection-key corruption),
  scalability (a limited number of hardware protection keys),
  and semantic difference (thread view versus process view).
\item \textbf{Software abstraction.}
  We design and implement \sys,
  a software abstraction to fully utilize \mpk.
  The protection key virtualization,
  metadata protection, and
  inter-thread key synchronization of \sys
  allow applications to effectively overcome
  the three challenges.
\item \textbf{Case studies.}
  We apply \sys to three applications,
  OpenSSL library, JavaScript JIT compiler, and Memcached,
  to show its effectiveness and practicality.
  \sys secures all of them
  with a few modifications and negligible overhead.
  \squishend

%
\PP{Organization.}
\autoref{s:mpk-analysis} explains
the current hardware and software supports of \mpk.
\autoref{s:challenge} describes
the limitations of \mpk.
\autoref{s:proposed} depicts
the design of \sys
to effectively resolve all the explained problems.
\autoref{s:app} shows
real-world applications 
that can benefit from \sys to improve their security.
\autoref{s:eval} evaluates
the security and performance characteristics of
\sys and the applications.
\autoref{s:discuss} discusses
the limitations of \sys and
possible approaches to overcome them.
\autoref{s:relwk} introduces related work and
\autoref{s:conclusion} concludes this paper.

\section{Intel MPK Explained}
\label{s:mpk-analysis}

In this section,
we describe the hardware design of Intel \mpk and current
kernel and library support.
Also, we check the performance characteristics of \mpk to show its efficiency.

\begin{figure}[t!]
  \centering
  \includegraphics[width=.95\columnwidth]{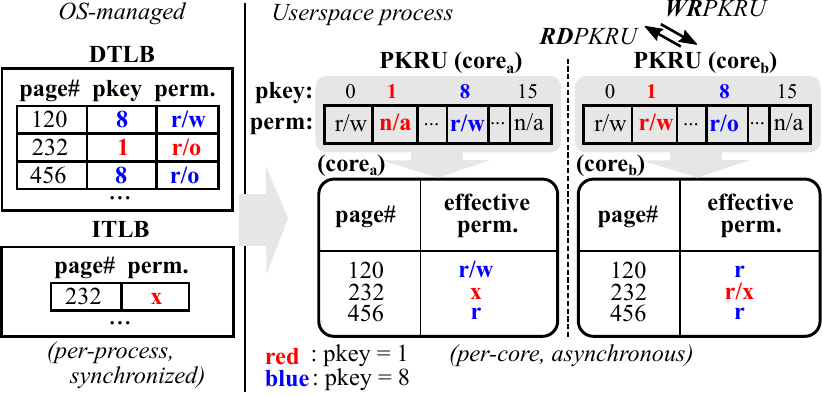}
  \caption{
    An example showing how MPK checks
    the permission of a logical core (hyperthread)
    on a specific memory page
    according to PKRU and page permissions.
    The intersection of the permissions
    determines whether a data access will be allowed.
    An instruction fetch is independent to the PKRU.
  }
  \label{f:mpk-check}
\end{figure}

\subsection{Hardware Primitives}
\label{ss:hardware}
Intel \mpk updates the permission of a group of pages
by associating a protection key to the group and
changing the access rights of the protection key instead of
individual memory pages
(\autoref{f:mpk-check}).
We explain the hardware primitives of \mpk.




\PP{Protection key field in page table entry}
%
%
%
\mpk assigns a unique protection key to a memory page group
to rapidly update its permission at the same time.
%
Intel CPUs with \mpk
utilize the previously unused four bits of each page table entry
(from 32nd to 35th bits)
to store a memory page's corresponding key value.
Thus, MPK supports up to 16 different page groups.
Since only supervised code can access and change PTEs,
the Linux kernel (from version 4.6) started to provide
a new system call, \pkmprotect,
to allow applications to
assign or change the keys of their memory pages
(\autoref{ss:kernel}).

\PPNoCheck{Protection key rights register (\pkru)}
A CPU with \mpk uses the value of \pkru
to determine its access right to each page group.
%
Two bits representing the right are
\emph{access disable} (AD) and
\emph{write disable} (WD) bits.
%
%
The value of $(AD,WD)$ represents
a thread's permission to a page group:
read/write $(0,0)$,
read-only $(0,1)$,
or no access $(1,x)$.
%
\pkru exists for each hyperthread
to provide a per-thread view.
%

%
%
%
%
%
%
%


\PP{Instruction set}
\mpk introduces two new instructions
to manage the \pkru:
(1) \wrpkru to update the protection information of
the \pkru and
(2) \rdpkru to retrieve the current protection information
from the \pkru.
\wrpkru uses three registers as input:
the \eax register containing new protection information
to overwrite the \pkru,
and the other two registers, \ecx and \edx,
filled with zeroes.
\rdpkru also uses the three registers for its operation:
it returns the current \pkru value via the \eax register
while overwriting the \edx register with 0.
The \ecx register also should be filled with 0
to execute \rdpkru correctly.
Note that, currently,
Intel does not document why \wrpkru and \rdpkru use
zeroed \ecx and \edx registers during their execution.

\subsection{Kernel Integration and Standard APIs}
\label{ss:kernel}
The Linux kernel supports \mpk since version 4.6,
and glibc supports \mpk since version 2.27.
%
They focus on
how to manage protection keys and
how to assign them to particular PTEs.
%
The Linux kernel provides three new system calls:
\pkmprotect, \pkalloc, and \pkfree.
It also changes the behavior of \mprotect
to provide execute-only memory.
Further, glibc provides two userspace functions,
\cc{pkey_get} and \cc{pkey_set},
to retrieve and update the access rights of a given protection key.
\autoref{t:standard-api} summarizes the APIs.



\begin{table}[t]
  \centering
  \footnotesize
  \resizebox{\columnwidth}{!}{
    \begin{tabular}{lrl}
  \toprule
   \textbf{Name} & \textbf{Cycles} &  \textbf{Description} \\
  \midrule
  \cc{pkey_alloc()}     & 186.3 & Allocate a new \cc{pkey} \\
  \cc{pkey_free()}      & 137.2 & Deallocate a \cc{pkey} \\
  \cc{pkey_mprotect()}  & 1,104.9 & Associate a \cc{pkey} key with memory pages\\
  \midrule
  \cc{pkey_get()/\cc{RDPKRU}} & 0.5 & Get the access right of a \cc{pkey}\\
  \cc{pkey_set()/\cc{WRPKRU}} & 23.3 & Update the access right of a \cc{pkey} \\
  \bottomrule
  \multicolumn{3}{r}{Ref. \cc{mprotect()}: 1,094.0 / MOVQ (\rbx to \rdx): 0.0 / MOVQ (\rdx to \xmm): 2.09} \\
\end{tabular}

  }
  \caption{Overhead of \mpk instruction,
    system calls, and standard library APIs.
    \emph{ref} shows the overhead of \mprotect and
    normal register move instructions for comparison.
    Each
    component is repeated 10 million times and the
    microbenchmarks are executed 10 times. }
  \label{t:standard-api}
\end{table}

\PP{pkey\_mprotect()}
The \pkmprotect system call
extends the \mprotect system call
to associate a protection key with
the PTEs of a specified memory region
while changing its page protection flag.
%
%
Interestingly,
\pkmprotect does not allow a user thread to
reset a protection key to zero,
the default protection key value
assigned to newly created memory pages.
We anticipate
this is in order to minimize the misuse of \mpk,
i.e., denying access to new pages
could result in an application crash.

\PP{pkey\_alloc() and pkey\_free()}
The Linux kernel provides
two other new system calls to
allocate and deallocate memory protection keys:
\pkalloc and \pkfree.
When a user thread invokes \pkalloc with access
right,
the kernel allocates and returns a protection key with
corresponding permission
according to
a 16-bit bitmap that tracks
which protection keys are allocated.
%
When a user thread invokes \pkfree,
the kernel simply marks the freed key as available in the bitmap.
The \pkmprotect function examines the bitmap afterward
to prohibit the use of non-allocated keys.
%


\PP{Execute-only memory}
The Linux kernel supports execute-only memory with \mpk.
If a user thread invokes \mprotect only with \cc{PROT_EXEC},
the kernel
(1) allocates a new protection key,
(2) disables the read and write permission of the key, and
(3) assigns the key to the given memory region.

\subsection{Quantifying Characteristics of Intel MPK}
\label{s:measure}

To evaluate the overhead and benefits of \mpk,
We measure
(1) the overhead of the \mpk instructions,
(2) the overhead of the \mpk system calls, and
(3) the overhead of \mprotect for contiguous memory and sparse memory.

\PP{Environment}
Our system consists of two Intel Xeon Gold 5115 CPUs (each CPU has 20 logical cores at 2.4GHz)
and 192GB of memory.
Linux kernel version 4.14 configured for \mpk is installed to this system.

\begin{figure}[t]
  \centering
  \footnotesize
  \input{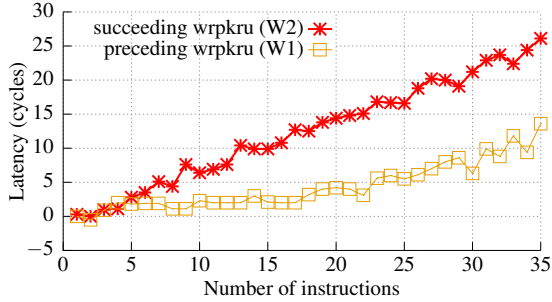}
  \caption{Effect of \wrpkru serialization on
    simple (i.e., ADD) instructions either
    preceding or succeeding \wrpkru (average of 10 million repetitions).}
  \label{f:flush}
\end{figure}

\PP{Instruction latency}
We measure the latency of \rdpkru and \wrpkru
to identify their micro-architectural characteristics.
\autoref{t:standard-api} summarizes the results.
The latency of \rdpkru is similar to
that of reading a general register, but
the latency of \wrpkru is high.
We anticipate that
\wrpkru performs serialization (e.g., pipeline flushing)
to avoid potential memory access violation due to
out-of-order execution.
To confirm this,
%
%
we insert a various number of \cc{ADD} instructions
before (\cc{W1}) and after (\cc{W2})
\wrpkru and
measure the overall latency (\autoref{f:flush}).
The results
show that \cc{W2} is always slower than \cc{W1},
implying that
the instructions executed right after
\wrpkru fail to
benefit from out-of-order execution
due to
the serialization.
%

%

\PP{System calls}
%
We measure the latency of the four Linux system calls for \mpk
(\autoref{t:standard-api}).
The latency of \mprotect and \pkmprotect on a 4\KB page is almost the same
because they use the same function \domprotect
internally.
%
%
\pkalloc and \pkfree are fast since they involve only simple
operations in the kernel, and the domain switching between
kernel and userspace dominates their time costs.

\begin{figure}[t!]
  \centering
  \input{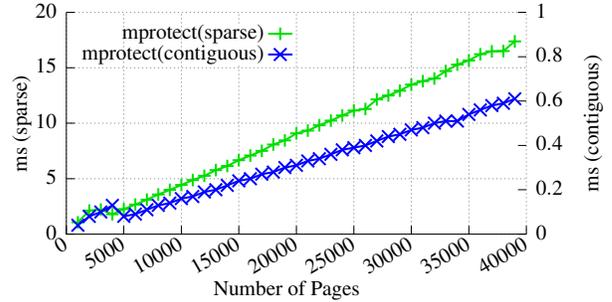}
  \caption{Overhead of \mprotect on contiguous and sparse
	memory (average cost of 10 million repetitions).
	Protecting contiguous pages takes shorter than protecting sparse pages.}
  \label{fig:shape}
\end{figure}

\PP{Contiguous versus sparse memory pages}
%
Using \mpk to change page permission only
involves an update on the \pkru and thus
is independent of the number of targeted pages and their
sparseness.
To show the performance benefit of \mpk over \mprotect,
we check how the number and
sparseness of the targeted pages affect the performance of
\mprotect.
To construct contiguous memory pages,
we call \mmap one time with certain memory size.
For sparse memory pages,
we call \mmap several times with one page size.
\autoref{fig:shape} shows that the overhead
of \mprotect increases in proportion to the number of pages.
This is because the number of pages affects how many VMAs
\mprotect needs to look up for permission
update.
Moreover,
the overhead of \mprotect becomes high when it is invoked on sparse memory pages
because we need to call \mprotect multiple times for each of them,
introducing frequent context switchings between kernel and
userspace.
%
%

\vspace{0.5em}
\boxbeg
\textbf{Summary.}
Intel \mpk allows a thread to rapidly change
the per-thread access rights to a group of pages
associated with the same protection key,
by updating a thread-local register \pkru
which only takes around 20 cycles.
Its performance is independent to
the number of pages composing a group and
their sparseness unlike \mprotect.
%
%
%
\boxend

\section{Challenges of Utilizing Intel MPK}
\label{s:challenge}
We studied Intel MPK can protect sensitive data
from unintended memory access
or expedite \mprotect for large memory (\autoref{s:mpk-analysis}).
However, a few challenges exist to utilize \mpk
in real-world applications, which are induced by either its inherent
features or the current software interfaces.
In this section,
We explain the challenges of using \mpk
in terms of security, scalability, and synchronization.

\subsection{Potential Security Problems}
\label{ss:threat}

Currently, \mpk suffers from two possible security problems 
occured by its interface design.

\PPNoCheck{Protection key use after free}
%
\pkfree just removes a protection key from a key bitmap and
does not update the corresponding PTEs.
Whether or not a key could already be associated with some pages,
the kernel will allocate the key if it is freed by \pkfree.
If a program obtains a key that is still associated with
some memory pages through \pkalloc,
the new page group will include unintended pages than it is supposed to.
A developer can face this vulnerable situation unconsciously
as current kernel implementation neither handles this automatically
nor checks if a free key is still associated with some pages.
The developer community also recognized the problem,
and recommends not to free the protection keys~\cite{glibc:pkey, kernel:pkey}.
Handling this problem superficially (i.e., wiping protection keys in PTE) 
without fundamental design change of
memory management in kernel will introduce huge performance overhead
because it requires to traverse the page table and VMAs to detect
entries associated with a freed key to update them, and flush all corresponding TLB entries.


\PP{Protection key corruption}
The existing OS supports
make the developer store
the allocated protection keys in the application's memory.
This design makes allocated protection keys
are vulnerable to corruption.
For example, after obtaining a key by invoking \pkalloc,
an application stores the key in its memory
to use the key later when it assigns
the key to more pages with \pkmprotect or
switches the permission using \wrpkru.
An attacker who managed to corrupt such keys
could manipulate one piece of code that has an access to
pages with one key into corrupting the pages with another key.
%
%



\subsection{Limited Hardware Resources}
\label{ss:hardwarelimit}
As Intel MPK relies on \pkru register for permission
change, it supports only 16 keys currently.
It is the responsibility of developers to ensure that
an application never creates more than 16 page groups at the same time.
%
This implies that
developers have to examine the number of active page
groups at runtime,
which are used by both the application
itself and the third-party libraries it depends on.
Otherwise, the program may fail to properly benefit from \mpk.
This issue undermines the usability of \mpk and discourages
developers from utilizing it actively.
%
%
Hardware solution like adding larger register (i.e., AVX) 
has limitation to scale because \mpk utilizes unused bits in PTE.
For example, 512 protection keys will demand 9 bits in page table and TLB entries, 
requiring enlarged entries or shrunk address bits.

\subsection{Semantic Differences}
To change the permission of any page group,
\mpk modifies the value of the \pkru.
%
However, the value is only effective in a single
thread because \pkru is thread-local
intrinsically as a register.
As a result, different threads in a process can have different
permissions for the same page group.
This thread-local inherence helps to
improve security for the applications that require isolation on
memory access among different threads,
but makes the semantic of \pkru change differ from that of \mprotect.
%
\mprotect semantically guarantees that page permissions
are synchronized among all threads in a process,
on which particular applications rely.
This not only makes it difficult to accelerate \mprotect with
\mpk, but also breaks the guarantee of
execute-only memory implemented on mprotect in latest kernel.
\mprotect supporting executable-only memory relied on \mpk 
does not consider synchronization among threads which 
developers basically expect to \mprotect.
Even when the kernel successfully allocates
a key for the execute-only page,
another thread might have a read access to it
due to a lack of synchronization.
%
To make \mpk a drop-in replacement of \mprotect for both security and usability,
developers need to synchronize the \pkru values
among all the threads.

\section{Software Abstraction of Intel MPK}
\label{s:proposed}

\begin{figure}[t!]
  \centering
  \includegraphics[width=.95\columnwidth]{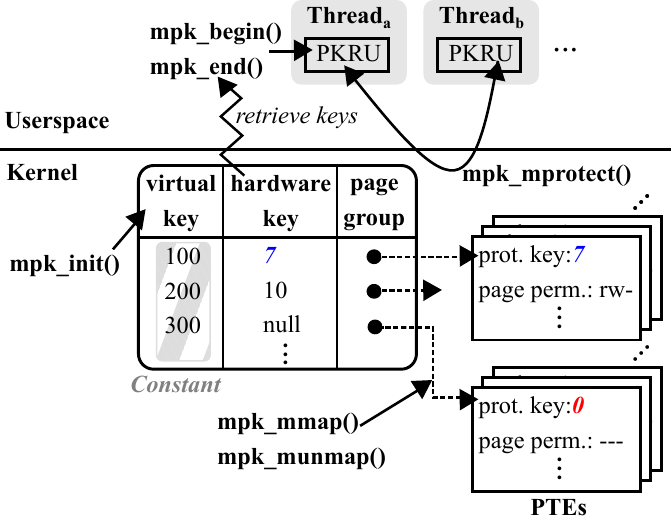}
  \caption{\sys overview. \mpkinit pre-allocates hardware keys
  and initializes the metadata table. \mpkmmap creates a page group
  with metadata,
  and \mpkdestroy destroys 
  the page group and the corresponding metadata.
  \mpkbegin and \mpkend provide domain-based thread-local isolation.
  \mpkprotect synchronizes permission changes globally.}
  \label{f:libmpk-overview}
\end{figure}

\sys provides a secure and usable abstraction for \mpk
by overcoming the challenges (\autoref{s:challenge}).
A developer can use \mpk easily
by either adding calls to \sys APIs
or replacing existing
\mprotect calls with those of \sys.
By decoupling the protection keys from APIs,
\sys is immunized against protection-key-use-after-free and
protection-key corruption.
Also, \sys allows an application to create more than
16 page groups by virtualizing the protection keys,
and provides a light-weight inter-thread \pkru synchronization mechanism.
%
\autoref{f:libmpk-overview} illustrates an overview of \sys.
The current version of \sys consists of 1.5K lines of code in total.

\PP{Threat model}
We assume an adversary who can corrupt non-control user data.
%
Arbitrary execution
is beyond our scope
because it allows the adversary to request system calls to directly manipulate permission.

\PP{Goals.} 
To utilize \mpk for domain-based isolation and as a substitute
for \mprotect,
we have to overcome the three challenges:
(1) security problems due to insecure key management,
(2) hardware resource limitations, and
(3) different semantics from \mprotect.
\sys adopts the three approaches (1) \emph{key virtualization}, (2)
\emph{metadata protection}, and (3) \emph{inter-thread key
synchronization} that effectively solve the challenges.

\begin{table}[t]
  \centering
  \footnotesize
   \resizebox{\columnwidth}{!}{
    \begin{tabular}{l@{~~}l@{~~}l}
  \toprule
   \textbf{Name} & \textbf{Argument} &  \textbf{Description} \\
  \midrule
    \cc{mpk_init} & \cc{evict_rate} & Initialize \sys with an eviction rate\\
    \cc{mpk_mmap} & \cc{vkey}, \cc{addr}, \cc{len}, \cc{prot} & Allocate a page group for a virtual key\\
                  & \cc{flags}, \cc{fd}, \cc{offset} & \\
    \cc{mpk_munmap} & \cc{vkey} & Unmap all pages related to a given virtual key\\
  \midrule
    \cc{mpk_begin} & \cc{vkey}, \cc{prot} & Obtain thread-local permission for a page group \\
    \cc{mpk_end} & \cc{vkey} & Release the permission for a page group \\
    \midrule
  \cc{mpk_mprotect} & \cc{vkey}, \cc{prot} & Change the permission for a page group globally \\
  \midrule
  \cc{mpk_malloc} & \cc{vkey}, \cc{size} & Allocate a memory chunk from a page group \\
  \cc{mpk_free} & \cc{size} & Free a memory chunk allocated by \mpkalloc\\
\bottomrule
\end{tabular}

   }
  \caption{\sys APIs.}
  \label{t:api}
\end{table}

\subsection{\sys API}
\label{ss:api}



\sys provides eight APIs shown in \autoref{t:api}.
To utilize \sys, an application first calls \mpkinit
to obtain all the hardware protection keys from the kernel
and initialize its metadata.
\mpkmmap allocates a page group for a virtual key, which should be a constant
integer that the developer passes.
%
%
%
\mpkdestroy destructs a page group by freeing a virtual key for
the group and unmaps all the pages.
\sys maintains the mappings between virtual keys and pages
to avoid scanning all pages at this destruction step.
On top of these, \sys also provides simple heap over
each page group (\mpkalloc and \mpkfree),
so that a developer can also use one or more page groups
to create a heap memory region for sensitive data.



\begin{figure}[t]
  \input{code/api.c}
  \coderule
  \caption{Example code for \sys APIs.}
  \label{c:api}
\end{figure}

\sys provides two usage models for developers.
The first model, a thread-local domain-based isolation model,
allows an application to temporarily unlock a page group
only for the calling thread.
\mpkbegin and \mpkend are the APIs for this model which
make a page group accessible and inaccessible, respectively.
The second model allows an application to quickly change
the access rights to a page group,
by replacing \mprotect with \mpkprotect.
%
\autoref{c:api} is example code
showing how to use \sys APIs.



\subsection{Protection Key Virtualization}
\sys enables an application to create more than 16 page groups
by virtualizing the hardware protection keys.
When an application creates a new page group by calling \mpkmmap,
a virtual key passed as argument is associated with new allocated
metadata for the new group.
%
%
The application uses the virtual key to
obtain or release the permission, or free the group,
while being prohibited from manipulating hardware keys.
%
The exact physical key that a page is associated with
is hidden from the program and a developer.

\begin{figure}[t!]
  \centering
  \begin{subfigure}[t]{\columnwidth}
    \centering
    \includegraphics[width=.85\textwidth]{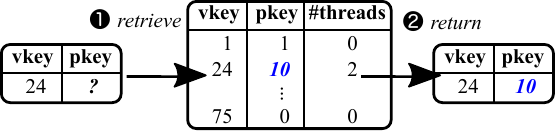}
    \caption{Hit case:
      \BC{1} A thread calls \protect\mpkbegin or \protect\mpkprotect with a vkey;
      \BC{2} \sys returns the corresponding pkey immediately.
    }
  \end{subfigure}
  \\
  \begin{subfigure}[t]{\columnwidth}
    \centering
    \includegraphics[width=.85\textwidth]{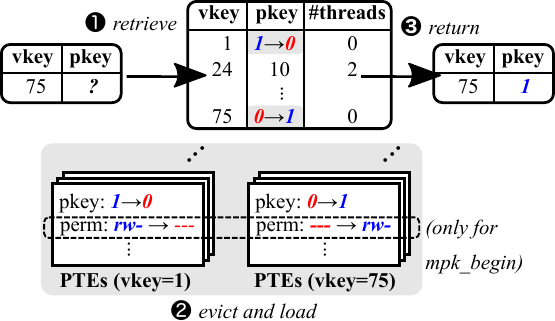}
    \caption{Miss case:
      \BC{1} A thread retrieves a vkey, but no corresponding pkey exists (pkey=0);
      \BC{2} \sys evicts the LRU pkey.
      In addition, \mpkbegin updates the page permission of the evicted and loaded page groups
      using \mprotect;
      \BC{3} \sys returns the new pkey.
    }
  \end{subfigure}
  \caption{
    Key virtualization in \sys. \emph{vkey} and
	\emph{pkey} represent a virtual key and its corresponding
	hardware protection key associated with a page group.
	\#threads indicates the number of threads
	running parallel inside a particular domain.}
  \label{f:key-virtualization}
\end{figure}

\sys handles mapping between virtual keys and hardware keys
like a cache (\autoref{f:key-virtualization}).
If a virtual key is already associated with a hardware key,
the virtual key exists inside the cache,
so further access with it produces a few latencies.
However, if the virtual key was not associated with a hardware key,
it needs to evict another virtual key or do nothing but just call
\mprotect for performance to change a permission.
The frequency of eviction or calling \mprotect is determined by
eviction rate.
%
%
The cache structure guarantees that a virtual key,
which changes a permission frequently, will be mapped
with a hardware key since it has a high possibility to be included inside the
cache.

\sys provides two policies to determine the mappings between
virtual and hardware keys.
%
%
When an application unlocks a page group thread-locally by
calling \mpkbegin, \sys always maps the group's virtual key with
a hardware key and uses it to grant access to
the \emph{calling thread}.
\sys maintains the mapping until the
thread calls \mpkend to release the access.
For this reason,
\sys cannot ensure that a calling thread
always obtains the access
due to hardware limitations.
%
That is,
if all hardware keys are actively used,
\sys is no longer able to provide any key.
In this case, \mpkbegin raises an exception and
lets the calling thread handle it
(e.g., sleeps until a key is available).
If a page group is not used by a thread,
\sys evicts the group by changing its protection key to 0 (default)
and revoking its page permission to disallow subsequent accesses.
Unlike \mpkprotect, \mpkbegin must evict a page group when every key is used 
to guarantee that a page group is permitted thread-locally.


The second policy, \mpkprotect,
also needs to map the virtual key to a hardware key,
but not exclusively.
%
Even when the page group is accessible,
\sys can unmap a hardware key and rely solely on the page attributes
because all threads have the access.
Hence, \sys maps only the page groups whose access rights
change frequently.
If \sys fails to find an available hardware key when it handles
\mpkprotect,
it unmaps and uses the least recently used (LRU) key for handling
\mpkprotect.
%
The hardware key of the evicted page group turns to 0.
To avoid excessive overhead due to frequent unmapping,
a developer can configure an \emph{eviction rate} to
control whether a hardware key has to be evicted according to
how frequent its permission updates.
In our approach, enforcing executable-only permission 
is not straightforward, 
because a conventional approach (i.e., \mprotect) 
does not support executable-only permission.
Therefore, \mpkprotect reserves one key for execute-only pages 
when an application creates them firstly, 
and does not evict this key until all executed-only pages disappear.
Every incoming executable-only permission request is guaranteed 
to get a hardware protection key 
to achieve executable-only permission.
If \mpkprotect already had executable-only page groups, 
further executable-only permission
requests will merge the incoming page groups 
with the existing executable-only page groups 
to utilize the reserved key.

\subsection{Metadata Protection}
\label{ss:metadata}

All metadata of \sys should not be corrupted by an attacker.
The first type of metadata is the virtual key
that an application uses to call \sys's API functions.
\sys assumes those virtual keys are hardcoded
in the application binary,
and the application never uses indirect calls
to access \sys's APIs.
\sys verifies it
by checking the binary at load time
to ensure that all direct invocations of \sys use hardcoded virtual keys
by checking the call site upon each invocation.

The second one is \sys's internal metadata:
the mappings between virtual keys and hardware keys,
and the page group information.
To protect these from malicious overwrites,
%
\sys maps one physical page into two virtual pages:
a read-only page for the application and a writable page for the kernel,
and stores the metadata in that page.
%
%
\sys slightly modifies existing system calls (e.g., \mmap, \munmap, and \mprotect)
to manage the metadata in the kernel.
Except metadata management, every management logic
for \sys is located in userspace to minimize unnecessary overhead from
domain switching between kernel and userspace.




\begin{figure}[t!]
  \centering
  \includegraphics[]{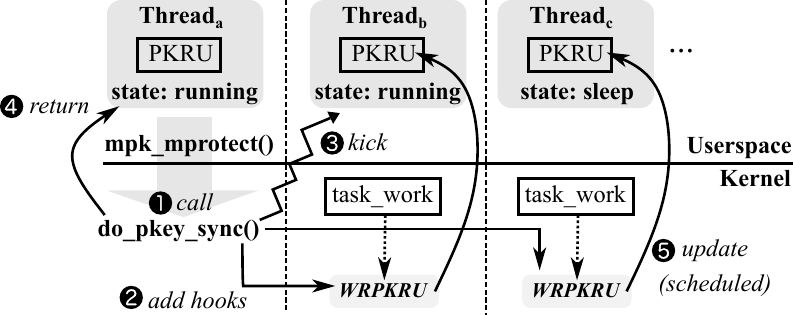}
  \caption{\pkru synchronization:
  \protect\circled{1} \mpkprotect calls \pkeysync to update
  the \pkru values of remote threads; \protect\circled{2}
  \pkeysync adds hooks to the threads' \cc{task\_work};
  \protect\circled{3} \pkeysync kicks all the running threads
  for synchronization; \protect\circled{4} \pkeysync returns to
  its caller; \protect\circled{5} The threads update their
  \cc{PKRU}s when they are scheduled to run.
}
  \label{f:sync0}

\end{figure}

\subsection{Inter-thread Key Synchronization}

\sys implements an inter-thread \pkru synchronization technique,
\pkeysync, for \mpkprotect for the two purposes:
(1) to ensure no thread has the read access to an execute-only
page and
(2) to replace existing page-table-based \mprotect for performance.
\pkeysync guarantees that a \pkru update is globally visible and effective
as soon as it returns.
Intuitively, this requires a synchronous inter-thread communication;
the calling thread needs to send messages to the other threads and wait until they
update the \pkru values and acknowledge it,
which suffers from a high cost.

We minimize the inter-thread \pkru synchronization latency
in a \emph{lazy} manner,
leveraging the fact that the \pkru values are utilized in the userspace.
%
If a remote thread is not currently being scheduled, it does not need to receive
the message immediately.
Even if the thread is currently being scheduled, it is enough for the thread to
have and update the \pkru values when it returns to the userspace.
If the calling thread can create a hook that the other threads will invoke right before
jumping back to the userspace and ensure that they are not in the userspace,
we can guarantee that all the other threads have the new \pkru values when
\pkeysync returns.

\autoref{f:sync0} illustrates the overall procedure of \mpkprotect.
\pkeysync utilizes an existing hooking point in the Linux kernel
to enforce the remote threads
to update the \pkru values right before returning to the
userspace and ensures that
all threads use the new \pkru values by sending rescheduling
interrupts.
In Linux,
a thread can have a list of callback functions (\cc{task_work})
that it will invoke at designated points,
including the return to the userspace.
A thread can register a callback for another by calling \cc{task\_work\_add()}.
In this way, \pkeysync lets the remote threads
update \pkru values lazily.
Although \pkeysync still needs to send inter-processor interrupts to ensure that no other
thread uses the old \pkru value after a certain point, our evaluation shows
that the overall latency of \mpkprotect is shorter than that of
\mprotect (\autoref{s:micro}).

\begin{table}[t]
  \footnotesize
  \centering
  \resizebox{\columnwidth}{!}{
    \begin{tabular}{l@{~~}l@{~~}l@{~~}l@{~~}l@{~~}l}
\toprule
  \textbf{Application} & \textbf{Protection} & \textbf{Protected data} &
  \textbf{\#pkeys} & \textbf{\#vkeys} & \textbf{Changed LoC} \\
\midrule
  OpenSSL & Isolation & Private
  key & 1 & 1 & 83\\
  JIT (key/page) & W$\oplus$X &
  Code cache & 15 & > 15 & CC 10 | SM 18 \\
  JIT (key/process) & W$\oplus$X &
  Code cache & 1 & 1 & CC 18 | SM 24 | v8 134 \\
  Memcached & Isolation & Slab,
  hashtable & 2 & 2 & 117\\
\bottomrule
\end{tabular}

  }
  \bigskip
  \caption{Three real-world applications of \sys. To enable
	W$\oplus$X in JavaScript engines, we use two approaches
	including using a virtual key for every page in the code
	cache, namely \emph{One key per page} and using a single
	protection key for all the pages in the code cache, namely
	\emph{One key per process}. CC and SM indicate Microsoft
  ChakraCore and Mozilla SpiderMonkey JS engine, respectively. 
  More code
	modification is required in Google v8 where W$\oplus$X is
	not originally supported. pkeys and vkeys mean protection 
  keys and virtual keys respectively}
  \label{t:apps}
\end{table}
\section{Applications}
\label{s:app}
We demonstrate the security benefit, efficiency, and
usability of \sys by augmenting three types of
popular applications:
an SSL library,
three JavaScript Just-in-time (JIT) compilation engines,
and an in-memory key-value store.
\autoref{t:apps}
summarizes the mechanisms
(e.g., page isolation or W$\oplus$X)
that we aim to provide
as well as
the protected data (e.g., key or code).
Evaluation results about these secure applications
are in \autoref{s:eval}.


\subsection{OpenSSL}
Transport layer security (TLS) or secure sockets layer (SSL) is
one of the most important security features to
prevent attacks from eavesdropping network traffic by encryption.
TLS/SSL relies on public-key cryptosystems,
such as RSA,
to authenticate communication parties and
to exchange a session key between them.
%
%
%

OpenSSL is a popular open-source library for offering encrypted
HTTPS and other secure communication by implementing the SSL and TLS protocols.
Although it is widely used open source project, it still suffer from diverse security
threat such as code execution, memory corruption or leak
sensitive information. Especially, information leak bugs are
powerful because a web server contains several sensitive data
(i.e., crypto key, password and personal private data).
%
For example, 
Heartbleed bug~\cite{heartbleed} is one of 
the information leak bug in OpenSSL, allowing
attacker to have a chance to leak sensitive information including private keys.


We apply \sys to OpenSSL to protect its private keys from
potential information leakage
by storing the keys in
isolated memory pages.
More specifically,
we first figure out all data types
that can store private keys (e.g., \cc{EVP_PKEY}) and
replace their heap memory allocation function
from \cc{OpenSSL_malloc()} to \mpkalloc
to store them in an isolated memory region.
Next,
we find all functions that need to access private keys
(e.g., \cc{pkey_rsa_decrypt()})
and let them access the isolated memory region
by inserting \mpkbegin and \mpkend before and after their call sites.
Note that
it is possible to
wrap individual, legitimate access to the isolated memory region
with \mpkbegin and \mpkend to minimize the attack window,
but it affects performance and programmability
such that this paper does not choose that approach.
%



\subsection{Just-in-time (JIT) Compilation}
\label{ss:jit}

JIT compilation
dynamically translates
interpreted script languages,
e.g., JavaScript and ActionScript,
into native machine code or bytecode
to avoid the overhead of
full compilation and repeated interpretation.
However,
it can suffer from security problems
because
it relies on writable code,
which potentially results in arbitrary execution.
To support JIT compilation,
the \emph{code cache} that
stores code generated at runtime needs to
be writable by a JIT compilation thread
to let it write and update compiled code,
and be executable by an execution thread.
This implies that
if attackers are able to compromise
the JIT compilation thread,
they make the execution thread
execute the code they provide.
%

ChakraCore~\cite{chakracore} and SpiderMonkey~\cite{spidermonkey}
mitigate the abovementioned problem
by enforcing the W$\oplus$X security policy on
the code cache with \mprotect.
They make the code cache writable
while disallowing execution
when they are updating code,
and, after it has updated,
they make the code cache executable
while disallowing write.
However,
they can suffer from \emph{race condition attacks}~\cite{song:sdcg15}
because they use \mprotect to change page permissions;
that is,
when a thread makes the code cache writable with \mprotect,
other threads
compromised by attackers
can also manipulate the code cache
with the same permission.

We apply \sys to the three popular JavaScript engines
(SpiderMonkey, ChakraCore, and v8)
to enforce the W$\oplus$X security policy
without the race condition problem
while ensuring better performance.
Unlike \mprotect,
\sys enables per-thread permission such that
it can ensure
only a JIT compilation thread has
a write right to the code cache.
%
%
We propose two approaches to implement the W$\oplus$X policy with \sys.

\PP{One key per page}
A context-free solution is replacing \mprotect with \sys
APIs to perform fast permission switches on targeted pages in
the code cache. All the protection keys are initialized with
read-only permission when a new thread is created. We dedicate
\emph{one protection key} to \emph{one page} when it is first
time re-protected via \mprotect, and change its page
permission to \cc{rwx}.  Later, we only need to call \mpkbegin
and \mpkend before and after when the JIT compiler updates the
corresponding page.  Based on the observation that mostly only
one page is updated at a time, we still invoke \mprotect if
multiple pages change permission.

\PP{One key per process}
We also propose a new approach specialized for JavaScript
engines, where only a single protection key is used to protect
all code cache. More specifically, when pages
are first time committed from the preserved memory region into
the code cache, they are assigned with the protection key and
their page permission is set to \cc{rwx}. Whenever any page in
the code cache is to be updated, the script engine needs
to call \mpkbegin and \mpkend.
Although more pages become temporarily writable, the
security of the code cache is ensured thanks to
the per-thread view of the protection key.



\subsection{In-Memory Key-Value Store}
In-memory key-value stores, such as Memcached,
are widely used to manage
a large amount of data in memory
to ensure low latency and high throughput.
%
Since the performance is
the most important requirement of it,
it usually does not adopt security techniques
that hinder its performance,
even if it stores sensitive information.
Especially,
security techniques whose performance
depends on the size of data
(e.g., \mprotect and encryption)
are avoided.
This implies that,
if an in-memory key-value store has
arbitrary read or write vulnerabilities,
attackers are able to
leak or corrupt sensitive information.

We apply \sys to an in-memory key-value store, Memcached,
to secure almost its entire memory.
\sys's performance is independent
to the size of memory to secure,
so it can efficiently work with
Memcached even when
the size of in-memory data is several gigabytes.
More specifically,
we secure
slabs that contain actual values and
hash tables that maintain key-value mappings
by replacing Memcached's \cc{malloc()} function
with \mpkalloc, and
let legitimate functions (e.g., \cc{ITEM_key()} and \cc{assoc_find()}) access them
by wrapping their call sites with \mpkbegin and \mpkend.
Note that,
in our current implementation,
we assign two different keys to
slabs and hash tables, respectively,
to narrow the attack surface.
It is possible to use more keys to
secure slabs in a fine-grained manner,
e.g., differentiating them according to their sizes.

\section{Evaluation}
\label{s:eval}
In this section,
we evaluate \sys in terms of its security
implication and performance by answering the following
questions:
\begin{itemize}
\item What security guarantees does \sys provide? (\autoref{s:esec})
\item Does \sys solve the security, scalability, and semantic-gap
  problems that existing \mpk APIs suffer from without
  introducing much performance overhead? (\autoref{s:micro})
\item Does \sys have negligible performance impact and
  outperform \mprotect in real-world applications? (\autoref{s:eapp})
\end{itemize}
The same system environment explained in \autoref{s:measure} is used
for performance evaluations.


\subsection{Security Evaluation}
\label{s:esec}

We first evaluate the security benefits from \sys regarding
memory protection and isolation.
%
%
For OpenSSL and Memcached, \sys provides domain-based
isolation to protect memory space that stores sensitive data.
The permission for the particular memory space set by \sys is
locally effective, which also prevents malicious accesses from
other compromised threads.
For example, \sys manages to prevent memory leakage from a
protected domain to outside. All attack attempts that exploit a
memory corruption vulnerability to leak or ruin sensitive data
stored in the isolated memory space are killed by segmentation
faults because they lack proper permission. To verify this, we
mimic the Heartbleed vulnerability by deliberately introducing a
heap-out-of-bounds read bug and inserting special
characters as a decoy private key placed next to the victim heap
region. When the vulnerability is triggered, OpenSSL hardened by
\sys crashes with invalid memory access.
However, \sys cannot fully mitigate memory leakage that
originates inside the protected domain. Thus, developers
should carefully design the domain to minimize the potential
attack surface when using \sys in their applications.

\sys can be used by JavaScript JIT compilers to guarantee
W$\oplus$X on the pages in the code cache. Unlike
\mprotect, \sys is immune to race condition attacks
launched by compromised threads running in parallel due to the
thread-local effectiveness of protection keys. When the JIT
compiler relies on \sys to switch the permission of a code page
for later updates, other threads controlled by attackers cannot
write malicious shellcode into the page simultaneously. To
verify it, we introduce two custom
JavaScript APIs for arbitrary memory read and write to
SpiderMonkey and ChakraCore, and test a simple PoC that
leverages these two APIs to locate the page of a function being
compiled and write shellcode to the corresponding page. Both
SpiderMonkey and ChakraCore crash with a segmentation fault at the
end.

\subsection{Microbenchmarks}
\label{s:micro}
We run several microbenchmarks to understand the performance
behavior of APIs in \sys.

\PP{Cache performance.}
\sys introduces cache to enable protection on more than 16 page
groups, whose performance is affected by its eviction rate and
hit rate, and the number of virtual keys in use.
We run the following two microbenchmarks to check
the cache performance.

\begin{figure}[t!]
  \centering
  \vspace{1.5em}
  ~~~~~~~~
  \input{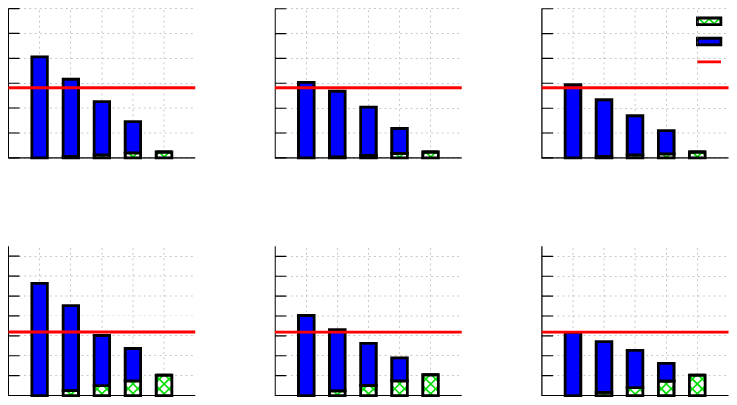}
  \vspace{-2.0em}
  \caption{Latency of \sys's key cache with
	various hit rates, eviction rates, and different number of
	threads. \mpkprotect and \mprotect are invoked on a 4\KB
	page.
	Red line marks the overhead of \mprotect. When the
	hit rate is 100\%, \mpkprotect is 12.2\X faster than
	\mprotect for one thread and 3.11\X faster for four threads.}
  \label{f:extension}
\end{figure}

\emph{Hit rate and eviction rate.}
The first benchmark measures cache performance with different
hit rates, eviction rates, and number of threads. We run the
benchmark with both one thread and four threads, where each
thread warms up by filling the key cache to evade cold miss and
invokes \mpkprotect on one page for a hundred times after 15
entries are filled.
\autoref{f:extension} presents the evaluation results, where
(1) the green box indicates the overhead incurred by the cache hit,
which is dominated by the time cost on \wrpkru and maintaining
internal data structures;
(2) the blue box indicates the overhead incurred by the cache miss,
which is dominated by the time cost on key eviction. More
specifically, \mpkprotect needs to unset the protection key that
is to be evicted and bind a new virtual key to it.
We test the microbenchmark with three eviction rates that indicate
the ratio of cache misses that eventually leads to key eviction.
If a cache miss occurs without key eviction, \mprotect is
invoked to change the permission of the pages.

Experimental results show that \mpkprotect outperforms
\mprotect except when the cache hit rate is
below 25\% with an eviction rate above 50\%.
This is because, unlike \mprotect, \mpkprotect
does not
merge and split the VMAs of targeted pages.
It becomes slow when being tested with four threads,
but is still comparable with \mprotect, whose latency
also increases in a multi-threading program.
%

\begin{figure}[!t]
  \centering
  \input{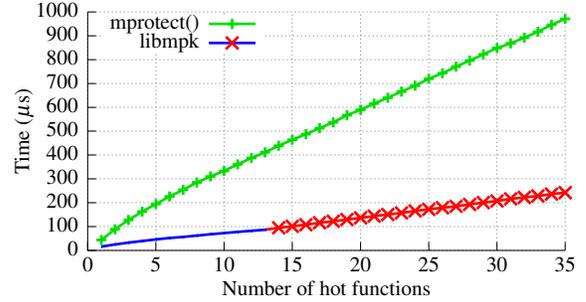}
  \vspace{1.0em}
  \caption{Average time cost to update permission
	when original and modified ChakraCore JIT-compile
	an increasing number of hot functions
    demanding distinct virtual keys.}
  \label{f:micro-js}
\end{figure}

\emph{Number of virtual keys.}
%
To evaluate how the number of used virtual keys affects the cache
performance of \sys, we re-implement W$\oplus$X in
ChakraCore in a \emph{one key per page} approach
(see \autoref{ss:jit}) and set the eviction rate as 100\%.
To introduce an increasing number of pages to be protected (\ie,
an increasing number of virtual keys to be used) during the
execution of ChakraCore, we design a simple microbenchmark.
The microbenchmark consists of a
set of JavaScript files from 1.js to N.js, each of which
contains N hot functions being invoked for 100,000 times
than its previous JavaScript file. For such a hot function,
ChakraCore allocates one more executable page to store the
native code and performs nine permission switches on the page
through one virtual key at runtime. Without any hot
function, ChakraCore allocates one page in the code
cache.
We run the original ChakraCore (version 1.9.0.0-beta) and the
modified one with our microbenchmarks (from 1.js to 35.js), and
record the time cost of changing permission of the pages in the
code cache (\ie, the execution time of \cc{VirtualProtect()} and
that of \mpkbegin and \mpkend) in total. Each JavaScript file is
executed 200 times, and the average time is presented in
\autoref{f:micro-js}.

The result shows that with \sys-based implementation of
W$\oplus$X, the time cost on permission switches linearly
increase when more hot functions are emitted and thus more
virtual keys are allocated to protect the code pages of the hot
functions. In particular, after 15 virtual keys are allocated
(marked in red), the time cost increases slightly faster than
before (marked in blue) due to cache eviction.
Nevertheless, the ChakraCore hardened by \sys still outperforms
3.2\X than the original ChakraCore using \mprotect to enforce
W$\oplus$X.

\PP{Memory overhead}
\sys dedicates memory space to store its internal data
structures for maintaining the metadata of these page groups
under protection (see \autoref{ss:metadata}). Each \mpkmmap
allocates 32 bytes of
memory to store the information of a new page group
(e.g., base address and size).
\sys maintains a hashmap to store the mapping between virtual
keys and hardware keys for fast query and access. In current
implementation, we pre-allocate 32\KB of memory for the hashmap,
and its size will automatically expand when a program
invokes \mpkmmap more than about 4,000 times.

\begin{figure}[t!]
  \centering
  \input{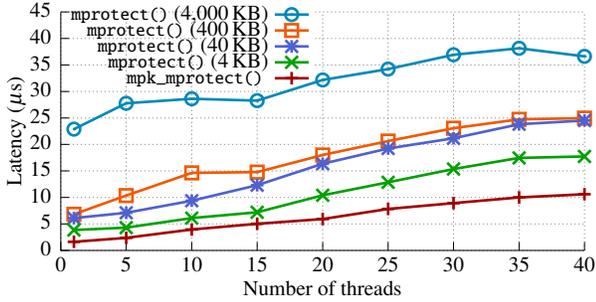}
  \vspace{1.0em}
  \caption{Latency of inter-thread permission synchronization
    using \mpkprotect and \mprotect calls on memory of varying sizes.
    \mpkprotect outperforms \mprotect 1.73\X for a single page
    and 3.77\X for 1,000 pages.}
  \label{f:sync}
\end{figure}

\PP{Synchronization latency}
\autoref{f:sync} shows the latency of
inter-thread permission synchronization
using \mpkprotect and \mprotect
on memory of varying sizes.
\mpkprotect is
1.73\X faster than \mprotect when
updating the permission of a single page.
The latency of \mprotect increases with the number of pages it
changes due to expensive operations on managing VMAs.
Compared to \mpkprotect, \mprotect costs at least 3.78\X
to change the permission of 1,000 pages.
%
%
The performance overhead of \mpkprotect
is independent of the number
of pages whose permission has been updated.
\autoref{f:sync} also shows that when there are many
threads, the latency of both \mprotect and \mpkprotect increases;
\mprotect flushes more TLBs, whereas \mpkprotect creates many
hooks in the kernel.

\subsection{Application Benchmarks}
\label{s:eapp}
We measure the performance overhead of \sys in practice by
evaluating three applications proposed in \autoref{s:app}.

\begin{figure}[t!]
  ~~~~~~~
  \input{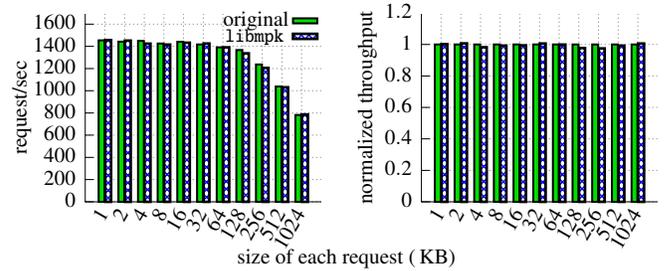}
  \vspace{-3.8em}
  \caption{Throughput of original \cc{httpd} and
	\cc{httpd} hardened by \sys.
    \sys slows down \cc{httpd} by at most 0.53\%.}
  \label{f:openssl}
\end{figure}

\PP{OpenSSL}
The Apache HTTP server~\cite{apache} (\cc{httpd})
uses OpenSSL to implement SSL/TLS protocols. To
evaluate the overhead caused by \sys, which is introduced to
protect private keys, we use ApacheBench to test
\cc{httpd} with both the original OpenSSL library and the modified one
with \sys. ApacheBench is launched 10 times and each
time sends 1,000 requests of different sizes from four concurrent
clients to the server.
We choose the DHE-RSA-AES256-GCM-SHA256 algorithm with 1024 bits key
as cipher suite in the evaluation.
\autoref{f:openssl} presents the evaluation result. On average,
\sys only introduces 0.58\% performance overhead in terms of the
throughput. The negligible overhead mainly comes from internal
data structure maintenance in \sys.

\begin{figure}[!t]
  \centering
  ~~~~~
  \resizebox{.97\columnwidth}{!}{
  \input{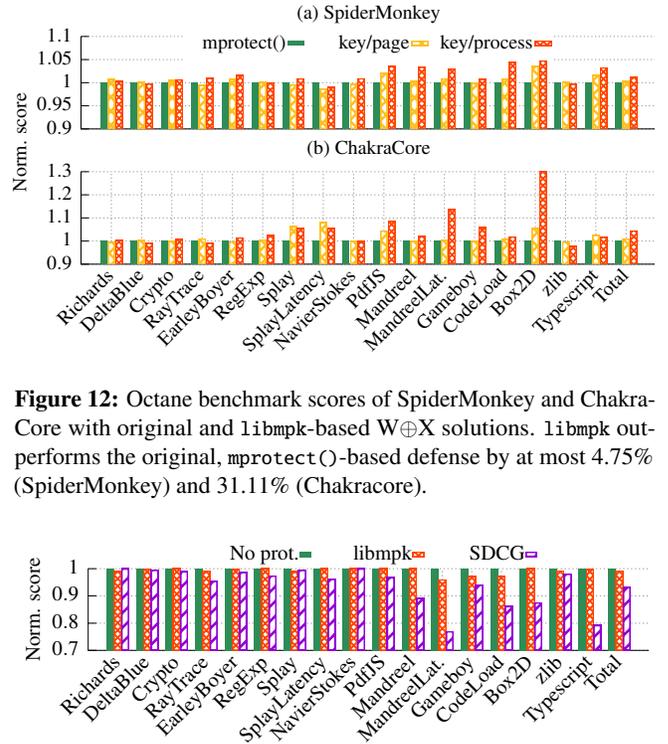}
}
  \vspace{1.5em}
  \caption{Octane benchmark scores of SpiderMonkey and
	ChakraCore with original and \sys-based
	W$\oplus$X solutions. \sys outperforms the original,
    \mprotect-based defense
    by at most 4.75\% (SpiderMonkey) and 31.11\% (Chakracore).}
  \label{f:firefox}
\end{figure}



\begin{figure}[!t]
  \centering
  \resizebox{1.05\columnwidth}{!}{
  \input{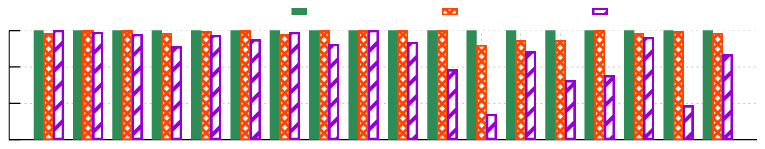}
}
  \vspace{0.1em}
  \caption{Octane benchmark scores of original v8 and two
	modified versions of v8 ensuring W$\oplus$X
	by SDCG and \sys. \sys only introduces 0.81\%
	overall performance overhead for W$\oplus$X in v8, compared
	with 6.68\% caused by SDCG.}
  \label{f:v8}
\end{figure}

\PP{Just-in-time compilation.}
We applied two proposed W$\oplus$X solutions based on \sys,
namely, \emph{one key per page} and \emph{one key per process}
(\autoref{ss:jit}) to both Spidermonkey (version 59.0) and
ChakraCore (version 1.9.0.0-beta) and evaluated their
performance with the Octane benchmark~\cite{octane} which
involves heavy JIT-compilation workloads at runtime.  Each
JavaScript program in the benchmark was directly executed by the
original and modified script engines for 20 times, and the
average score was recorded. Figure~\ref{f:firefox} shows the
final results.

For SpiderMonkey, both \sys-based approaches
outperform the \mprotect-based approach on the total score,
namely, 0.38\% and 1.26\%, which is consistent with the claim
from Firefox developers that enabling W$\oplus$X with \mprotect
in SpiderMonkey introduces less than 1\% overhead for the Octane
benchmark.
The reason is that SpiderMonkey is designed to get rid of
unnecessary \mprotect calls when its JIT compiler works.
The performance scores of nearly all the programs increase
through \emph{on key per page} (at most 3.60\% on \cc{Box2D})
and \emph{one key per process} (at most 4.75\% on \cc{Box2D}),
except for \cc{SplayLatency} protected by \emph{one key per
page}, whose score is dropped by 1.36\%.
%
%
%
When a large amount of new executable pages are allocated at
runtime but updated few times afterward, the script engine fails
to benefit from fast permission switches through \wrpkru, but
suffers from intensive cache eviction when \emph{one key per
page} is applied.

Our two \sys-based approaches improve ChakraCore by 1.01\% and
4.39\% on the total score of the Octane benchmark,
respectively.
ChakraCore is suitable for \sys-based W$\oplus$X solutions since it only
makes one page writable per time regardless of
emitted code size.
%
Note that \emph{one key per page} increases the performance
score of ChakraCore at most 7.96\% when testing
\cc{SplayLatency} while \emph{one key per process} improves the
performance by mostly 31.11\% on \cc{Box2D}.
%
Nevertheless, the score of \cc{zlib} decreases by 2.12\% when
\emph{one key per process} is applied.
This is because when new executable pages are committed, they
are protected with the single protection key, which requires an
extra invoking of \pkmprotect on multiple pages. If these pages
are rarely updated afterward, the introduced \pkmprotect calls
hurt the overall performance.

The \mprotect-based approach is vulnerable to the race condition
attack figured out by SDCG~\cite{song:sdcg15} (see
\autoref{ss:jit}). SDCG protects the JIT code pages of v8 with
W$\oplus$X by emitting the code in a dedicated process. The code
pages are not writable in other processes, which prevents the
attack.
To demonstrate the performance advantage of our in-process
\sys-based approaches, which are also free of race condition
attacks, we applied one of our approaches, \emph{one key per
process}, to Google v8 (version 3.20.17.1 used
in~\cite{song:sdcg15}) and evaluated the performance through the
Octane benchmark as well. Figure~\ref{f:v8} presents the
performance comparison among original v8, v8 with SDCG, and v8
with \sys. Note that originally, v8 has not deployed W$\oplus$X
to protect its code cache so far. Our approach only introduces
0.81\% overall performance loss, compared with 6.68\% caused by
SDCG.

To summarize, our \sys-based approaches, which are free of the
race condition attack, outperform the \mprotect-based approach
currently applied in practice to enforce W$\oplus$X protection
on code cache pages with negligible overhead.

\begin{figure}[t]
  ~~~~~~
  \input{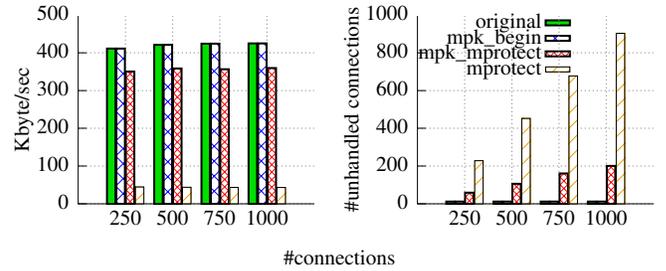}
  \vspace{-4.5em}
  \caption{Throughput and unhandled concurrent connections
	of original Memcached and three versions of Memcached whose
	key-value pairs are protected by \mpkbegin, \mpkprotect, and
	\mprotect. \mpkbegin's overhead is negligible
    in comparison to the original. \mpkprotect outperforms
    \mprotect 8.1\X while ensuring same semantics.}
  \label{f:memcached}
\end{figure}

\PP{In-memory key-value store.}
\label{sss:inmemoryeval}
To study the performance overhead of \sys when protecting large
memory, we evaluate the modified Memcached whose
key-value pairs are isolated by \sys. More specifically,
the modified Memcached pre-allocates 1\GB memory,
which is used instead of slab pages allocated by glibc
\cc{malloc()} to store key-value pairs. Besides the original
Memcached, we also evaluate the Memcached whose key-value pairs
are protected by \mprotect. To study the performance
of \mpkprotect in real-world applications, we also create
the Memcached guarded by \sys with permission synchronized as
another evaluation target for comparison. Each aforementioned
version of Memcached launches with four concurrent threads, and we
connect to it remotely through \cc{twemperf}~\cite{twemperf}.
We create from 250 to 1,000 connections per second, and 10
requests are sent during each connection.

\autoref{f:memcached} presents the evaluation results. The
modified Memcached hardened by \sys only has 0.01\% overhead in
terms of data throughput and almost no overhead regarding
concurrent connections processed per second, which indicates
that \sys performs well even when protecting a huge number of
pages.
By contrast, \mprotect introduces nearly 89.56\%
overhead in terms of data throughput when protecting 1\GB memory
in Memcached and a large number of unhandled concurrent connections
accumulate in this case. This is because \mprotect
involves page table traversing, which is considered expensive
when dealing with a large number of pages.
To evaluate the synchronization service of \sys in practice, we
also run Memcached protected by \mpkprotect.
This design ensures the same
semantics but outperforms \mprotect 8.1\X regarding
throughput.

\sys provides the same functionality of \mprotect
with much better performance when protecting huge size memory.
Moreover, in multi-threading applications, using \mprotect to
ensure in-thread memory isolation requires lock, which is not
required when using \sys due to its inherent property.

\section{Discussion}
\label{s:discuss}


We discuss potential attacks
on both Intel \mpk and \sys.

\PP{Rogue data cache load (Meltdown)}
We found Intel \mpk can suffer from
the rogue data cache load,
also known as the Meltdown attack~\cite{meltdown, intel-meltdown-spectre}.
The rogue data cache load is possible
because
current Intel CPUs check the access permission to
a specific memory page
after they have loaded it into the cache.
\mpk is not an exception because
Intel CPUs check the access rights of \pkru
when checking the page permission
at the same pipeline phase.
This allows attackers to infer
the content of a present (accessible) page
even when its protection key has
no access right.
Since Intel is considering
hardware-level mitigation techniques against
the rogue data cache load~\cite{intel-meltdown-spectre},
we believe this problem will be solved in the near future.

\PP{Control flow hijacking}
Developers call \pkmprotect to set a protection key for a group
of pages followed by a series of \cc{WRPKRU}s to change the
permission of the page group.
These two default interfaces provide a new attack surface when
the control flow of a compromised process has been hijacked.
More specifically, attackers can change the \pkru through
\wrpkru to get permission for accessing an isolated memory space.
The protection key for specific memory space can also be changed
by attackers through \pkmprotect.
To provide firm protection on isolated memory space, developers
can adopt sandboxing techniques~\cite{ford:vx08, yee:native09,
li:minibox14} to prevent attackers from invoking these two
operations and harden the control flow of their
applications~\cite{kuznetsov:cpi14, zhang:vtint15, abadi:cfi05,
koo:ccrsp18}.




\section{Related Work}
\label{s:relwk}

 \sys abstracts \mpk, which can primarily be used
 for memory isolation in the context of security.
 Though only the latest Intel processors provide \mpk,
 other architectures also support
 similar hardware features for grouping pages,
 as discussed in~\autoref{s:intro}.
%
%
 We introduce proposed applications of
 \mpk and ARM Domain,
 and other memory isolation mechanisms
 designed for different applications.
 Note that some of them have the threat models
 that are different from ours.
\PP{\mpk applications}
During conducting our study,
we noticed that 
there were a few ongoing studies using
MPK to achieve different security goals.
Burow et al.~\cite{burow:shining} leverage 
both MPK and memory protection extension (MPX) 
to efficiently isolate the shadow stack.
ERIM~\cite{erim:vahldiek} utilizes MPK
to isolate sensitive code and data.
In addition, XOM-Switch~\cite{zhang:xomswitch18} 
relies on MPK to enable execute-only memory 
for unmodified binaries.
%
%
Our effort on providing a software abstraction for \mpk is
orthogonal to these studies, which are all potential
applications of \sys.
%
%
These schemes can leverage 
\sys to achieve secure and scalable key management to create as many
sensitive memory regions as required securely.

\PP{Memory protection with similar hardware features}
Memory protection mechanisms have been 
leveraging new hardware features for efficiency
(e.g., software-based fault isolation~\cite{wahbe:sfi,sehr:adapting}).
ARM has a hardware feature named as \textit{Domain}~\cite{arm} 
and IBM Power supports \textit{Storage Protection}~\cite{power}.
ARMlock~\cite{armlock:zhou}, FlexDroid~\cite{seo:flexdroid}, 
and Shreds~\cite{chen:shreds16} rely on ARM Domain~\cite{arm}
to isolate untrusted program modules, 
third-party libraries, and sensitive code modules,
respectively.

Although they have a similar high-level concept compared to \mpk,
their underlying designs make them have different low-level
behaviors and potential benefits.
Domain differs from \mpk in two folds: 
the way of defining and switching permissions.
To change the permission for one or more page groups using Domain,
a thread updates a register called \textit{Domain Access Control Register} 
(DACR) which defines a running thread's access rights 
to a particular page group.
Unlike \mpk, Domain does not allow an application to update
the register by itself.
Instead,
it requires the application to invoke a system call 
since only the OS kernel can update DACR.
Furthermore, Domain does not support execute-only pages 
because it does not allow a thread to 
define an \textit{additional} access rights 
for a page group.
To make a page execute-only using \mpk, 
a program can make the page executable
through \mprotect and forbid page access by using \mpk.
By contrast, if a program creates non-readable 
pages through Domain, the processor cannot fetch any
instructions from that page.
IBM Storage Protection allows a program to create 32 different page groups,
and uses two special registers to determine the permissions on
every page group a running thread owns as \mpk does.
%
%
%
Similar to \mpk, Storage Protection has a restriction 
on the number of page groups under control.
Moreover, there does not exist any software abstraction 
to overcome this limitation.
Nevertheless, Storage Protection 
provides protection keys 
for kernel memory space unlike \mpk.
%
%

\PP{Software-based fault isolation (SFI)}
SFI~\cite{wahbe:sfi} prohibits 
unintended memory accesses
by inserting address masking instructions 
just before load and store instructions.
%
%
Sandboxing mechanisms, 
such as Native Client (NaCl)~\cite{sehr:adapting,nacl:chrome},
relies on SFI to isolate untrusted code.
Code-Pointer Integrity~\cite{kuznetsov:cpi14} also uses SFI
to protect the code pointers from unsanitized memory accesses.
Also, MemSentry~\cite{koning:no17} provides 
a unified memory isolation framework
based on hardware features
to reduce the performance overhead of SFI.
SFI enables an application to partition its memory into multiple regions,
but the cost of address masking limits the shape of partitions,
which are commonly contiguous pieces of memory.
By contrast, \mpk enables an application to 
partition the memory into the regions with arbitrary shape.
Further, the overhead of SFI on address masking 
increases by the number of isolated memory regions 
unlike \mpk.

\PP{Multiple virtual address space}
%
Using multiple virtual address spaces (i.e., multiple page tables) for a program
can protect the memory of sensitive or untrusted components from the others.
SMV~\cite{hsu:smv16} uses multiple page tables
to isolate the memory of threads in a single process from each other,
which is similar with \cite{privtrans:brumley,provos2003preventing,wang:arbiter15}.
Other systems~\cite{litton:lwc16,bittau:wedge08,hajj:spacejmp16}
also provide different memory views to individual threads or small execution units
using separated page tables.
Kenali~\cite{song:kenali} 
uses a page-table-based isolation mechanism
to protect sensitive data in which a separate page table is
created for each thread.
%
Unlike \sys,
these mechanisms suffer from non-negligible performance
overhead due to slow and frequent page table switches.

\section{Conclusion}
\label{s:conclusion}

Intel \mpk supports efficient per-thread permission control on groups of
pages. However, \mpk's current hardware implementation and software interfaces
suffer from security, scalability, and
semantic-gap problems.
To solve these problems,
\sys proposes a secure, scalable, and semantic-gap-mitigated
software abstraction of \mpk for developers to perform fast memory
protection and domain-based isolation in their applications.
Evaluation results show that \sys incurs negligible
performance overhead (<1\%)
for domain-based isolation and better
performance for a substitute of \mprotect when
adopted to real-world
applications: OpenSSL, JavaScript JIT compiler, and Memcached.



\bibliographystyle{acm}
{\footnotesize
\bibliography{p,sslab,conf}}

\end{document}